\documentclass[aip,apl,reprint,superscriptaddress]{revtex4-1}

\usepackage{graphicx}  
\usepackage{dcolumn}   
\usepackage{bm}        
\usepackage{amssymb}   
\usepackage{amsmath}
\usepackage{hyperref}
\usepackage{color}
\usepackage{multirow}
\usepackage{ifpdf}
\usepackage{tikz}
\usetikzlibrary{shapes,calc}
\usepackage[normalem]{ulem}
\usepackage{courier}

\newcommand{\SM}[0]{{Supplemental Material}}
\renewcommand{\vec}[1]{\mathbf{#1}}
\renewcommand{\t}[1]{\text{#1}}

\newcommand{\exciting}{{\usefont{T1}{pcr}{n}{n}exciting}}
\newcommand{\FHIaims}{{\usefont{T1}{pcr}{n}{n}FHI-aims}}
\newcommand{\GPAW}{{\usefont{T1}{pcr}{n}{n}GPAW}}
\newcommand{\VASP}{{\usefont{T1}{pcr}{n}{n}VASP}}

\begin{document}

\title{Numerical Quality Control for DFT--based Materials Databases}

\author{Christian Carbogno}
\affiliation{Fritz-Haber-Institut der Max-Planck-Gesellschaft, Faradayweg 4--6, D-14195 Berlin, Germany}
\author{Kristian Sommer Thygesen}
\affiliation{Center for Atomic-scale Materials Design (CAMD), Department of Physics, Technical University of Denmark, Fysikvej 1 2800 Kgs. Lyngby, Denmark}
\author{Bj\"orn Bieniek}
\affiliation{Fritz-Haber-Institut der Max-Planck-Gesellschaft, Faradayweg 4--6, D-14195 Berlin, Germany}
\author{Claudia Draxl}
\affiliation{Physics Department and IRIS Adlershof, Humboldt-Universit\"at zu Berlin, Zum Gro\ss en Windkanal 6, D-12489 Berlin, Germany}
\affiliation{Fritz-Haber-Institut der Max-Planck-Gesellschaft, Faradayweg 4--6, D-14195 Berlin, Germany}
\author{Luca M. Ghiringhelli}
\affiliation{Fritz-Haber-Institut der Max-Planck-Gesellschaft, Faradayweg 4--6, D-14195 Berlin, Germany}
\author{Andris Gulans}
\affiliation{Physics Department and IRIS Adlershof, Humboldt-Universit\"at zu Berlin, Zum Gro\ss en Windkanal 6, D-12489 Berlin, Germany}
\author{Oliver T. Hofmann}
\affiliation{Institute of Solid State Physics, Graz University of Technology, NAWI Graz, Petergasse 16, 8010 Graz, Austria}
\author{Karsten W. Jacobsen}
\affiliation{Center for Atomic-scale Materials Design (CAMD), Department of Physics, Technical University of Denmark, Fysikvej 1 2800 Kgs. Lyngby, Denmark}
\author{Sven Lubeck}
\affiliation{Physics Department and IRIS Adlershof, Humboldt-Universit\"at zu Berlin, Zum Gro\ss en Windkanal 6, D-12489 Berlin, Germany}
\author{Jens J\o{}rgen Mortensen}
\affiliation{Center for Atomic-scale Materials Design (CAMD), Department of Physics, Technical University of Denmark, Fysikvej 1 2800 Kgs. Lyngby, Denmark}
\author{Mikkel Strange}
\affiliation{Center for Atomic-scale Materials Design (CAMD), Department of Physics, Technical University of Denmark, Fysikvej 1 2800 Kgs. Lyngby, Denmark}
\author{Elisabeth Wruss}
\affiliation{Institute of Solid State Physics, Graz University of Technology, NAWI Graz, Petergasse 16, 8010 Graz, Austria}
\author{Matthias Scheffler}
\affiliation{Fritz-Haber-Institut der Max-Planck-Gesellschaft, Faradayweg 4--6, D-14195 Berlin, Germany}

\date{\today}

\begin{abstract}
Electronic-structure theory is a strong pillar of materials science. Many different computer codes that employ different approaches are used by the community to solve various scientific problems. Still, the precision of different packages has only recently been scrutinized thoroughly, focusing on a specific task, namely selecting a popular density functional, and using unusually high, extremely precise numerical settings for investigating 71 monoatomic crystals\cite{Cottenier16}. Little is known, however, about method- and code-specific uncertainties that arise under numerical settings that are commonly used in practice. We shed light on this issue by  investigating the deviations in total and relative energies as  a function of computational parameters. Using typical settings for basis sets and $\mathbf{k}$-grids,  we compare results for 71 elemental\cite{Cottenier16} and 63 binary solids obtained by three different electronic-structure codes that employ fundamentally different strategies. On the basis of the observed trends, we propose a simple, analytical model for the estimation of the errors associated with the basis-set incompleteness. We cross-validate this model using ternary systems obtained from the NOMAD Repository and discuss how our approach enables the comparison of the heterogeneous data present in computational materials databases.
\end{abstract}

\maketitle

Over the last decades, computational materials science has evolved as a paradigm of materials science, complementing theory and experiment with {\it computer experiments}.\cite{Handbook} In particular, density-functional theory (DFT) has become the workhorse for a plenitude of computational investigations, representing a good compromise between precision and computational expense, thus allowing for the investigation of realistic systems with affordable numerical effort.\cite{Jones15} The widespread application of electronic-structure theory was especially fueled by the development and distribution of many user-friendly and computationally efficient simulation packages (termed {\it codes} in the following) based on DFT. Essentially all these codes rely on the same fundamental physical  concept and solve the Kohn-Sham (KS) equations\cite{Kohn65} of DFT self-consistently by expanding the Kohn-Sham states in a finite basis set.  Moreover, apart from the choice of the basis set, different approximations and various numerical techniques and algorithms are employed. Inherently, this raises the question how consistent, and hence, how comparable, results from different codes are.

Only recently, a synergistic community effort led by  K. Lejaeghere and S. Cottenier\cite{Cottenier16} has shed light on these issues,  essentially concluding that ``most recent codes and methods converge 
toward a single value''.  This concerns, however, only the investigated relatively robust case of computing the equation of states for elemental solids\cite{Lejaeghere14,Cottenier16} using the PBE exchange-correlation (xc) functional. In this context, it has to be noted that such a close agreement across codes and methods  was only achieved by using  {\it safe} numerical settings that  guaranteed highest precision and that are rarely used in routine DFT calculations. In practice, such  settings are often not even necessary  as long as only data obtained by the same methodology, code, and settings are used, because then one benefits from error cancellation, and trends are described reliably. 

Over the last decade, the increased amount of available computational power as well as the maturity of existing first-principles  materials-science codes made it possible to perform computational studies in a ``high-throughput'' fashion by scanning   the compositional and structural space in an almost automated manner.\cite{Curtarolo2013,Gaultois2013,Saal:2013ht,Hachmann2014} In  such a case, the numerical settings have to be chosen \textit{a priori} in such a way that the trends of the properties of interest are captured. Often, this is achieved via educated guesses, sometimes via (semi-)automatic algorithms.\cite{Jain2015,Pizzi2016} Since the properties of interest differ in different investigations, also the numerical settings can vary quite significantly.\cite{Fischer06,Wang12,Castelli2012}  This has some impact on the possibility of reusing data beyond its original scope and purpose. Also, comparing data from different sources -- created using different methodologies and settings or focusing on different properties -- is not risk-free, in spite of the fact that the data may be publicly available in databases and repositories, as for instance, in the NOMAD Repository,\cite{Draxl:2018ec} AFLOW,\cite{curtarolo:art104,curtarolo:art75} Materials Project,\cite{Jain2013} OQMD,\cite{Saal:2013ht} Materials Cloud,\cite{Talirz:2020ve} the Computational Materials Repository,\cite{Haastrup:2018ca}  and alike. In a nutshell, using data from different sources that are based on different numerical settings implies potentially uncontrollable  uncertainties. This is a pressing and severe issue, given that the sheer amount of calculations existing to date prevents a human, case-by-case check of the data. 

In this work, we describe a first step for overcoming this unsatisfactory situation and show how errors for data stemming from  DFT computations can be estimated. We emphasize that we do not investigate errors that originate from the use of approximate {\it physical} equations,~e.g.,~the use of a  particular xc-functional.  We rather focus on {\it numerical} aspects,~i.e.,~on errors arising  from the fact that the same equation is solved in different approaches by employing 
different numerical approximations and 
techniques.\footnote{Different treatments of exchange and correlation can, however, require different numerical settings for convergence, as discussed in Sec.~\ref{Outlook}.} 
To this end, we systematically investigate the numerical errors that arise in total energies and energy differences when three different methodologies are applied,  using representative DFT codes as examples.
These are the \textit{linearized augmented plane-waves plus local orbitals} ansatz, as implemented in the  all-electron, full-potential code \exciting\cite{Gulans2014exciting}, the \textit{linear combination of numeric atom-centered orbitals~(NAOs)}
method as implemented in the  all-electron, full-potential code \FHIaims\cite{Blum09,Havu:2009ug}, as well as the \textit{projector-augmented wave~(PAW)} formalism\cite{blochl_1994} implemented in the package \GPAW.\cite{gpaw2005,gpaw2010} All electrons are accounted for  on the same footing in the self-consistency cycle in the first two methods. Conversely, core states are frozen in the PAW approach and valence states are mapped onto smooth pseudo-valence states using a linear transformation involving atom-centered partial wave expansions.\cite{blochl_1994} These pseudo-states are smooth and represented in a plane-wave expansion.\footnote{ Throughout this work, we use the PAW potentials recommended by the \GPAW\ developers.} In the following, we evaluate and analyze the numerical errors arising in these different formalisms at various levels of precision and then suggest how to estimate the errors associated with the basis-set incompleteness and, consequently, get access to the complete-basis-set limit for total energies and energy differences. 

\begin{figure*}
  \centering
  \includegraphics[clip,width=0.95\linewidth]{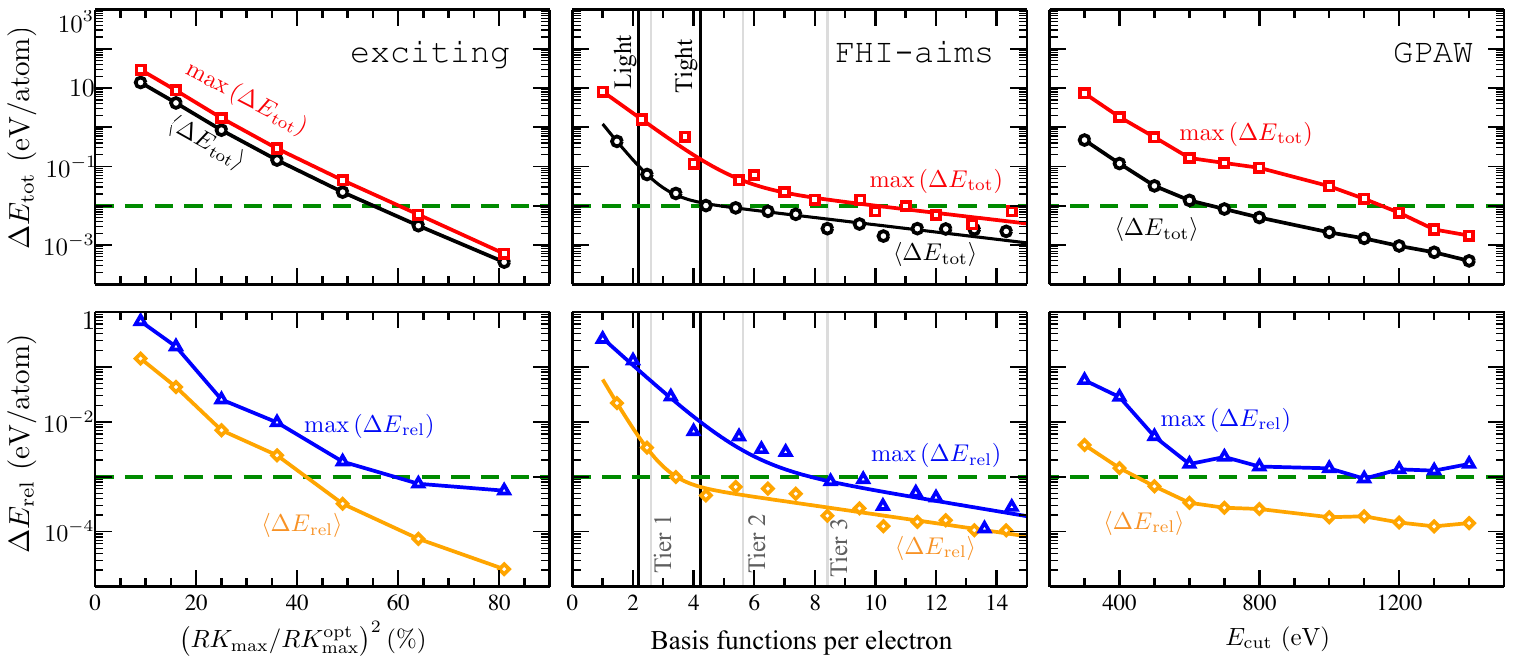}
  \caption{Average  (black/orange) and maximum  (red/blue) error in $E_\t{tot}$~(upper row) and $E_\t{rel}$~(lower row) as a function of basis-set size as observed in \exciting, \FHIaims, and \GPAW\ for the 71 elemental solids (PBE, 8$\mbox{\AA}$~$\vec{k}$-point density). Note that we use a logarithmic scale for the energy axis and different energy windows in the upper and lower row. See text for a discussion of the  different metrics used in the different codes to quantify basis set quality. The green dashed horizontal lines indicate values of typically wanted precision.  
} 
\label{fig:error_basis}
\end{figure*}

\section{Methods}
To perform the DFT calculations with these three codes in a systematic manner, the atomic simulation environment ASE\cite{Bahn2002,ase2017} was used 
to generate the code-specific input files
and to store the results using ASE's lightweight database module.
In this paper, we focus on the two main numerical approximations that are used to 
discretize and represent the electron density  $n(\vec{r})=\sum_{l{\bf k}}|\psi_{l{\bf k}}(\vec{r})|^2$ via the Kohn-Sham 
wavefunctions~$\psi_{l{\bf k}}(\vec{r})$ for the individual electronic states~$l$. 
These are the density of the reciprocal-space grid ({\bf k} grid) for Brillouin-zone (BZ) integrations and the finite 
basis set~$\phi_{j{\bf k}}(\vec{r})$ enumerated via~$j$. 
The Kohn-Sham wavefunctions are written as
\begin{equation}
\psi_{l\vec{k}}(\vec{r})= u_{l\vec{k}}(\vec{r}) \exp(i\vec{k}\vec{r}) \quad \text{with} \quad u_{l\vec{k}}(\vec{r}) = \sum_j c_{lj{\bf k}} \phi_{j{\bf k}}(\vec{r}) \;.
\end{equation}
For the BZ sampling, we use a $\Gamma$-centered $\vec{k}$-grid characterized by a uniform $\vec{k}$-point density
\begin{equation}
\rho_{\vec{k}}=(N_{\vec{k}}/V_\t{BZ})^{\frac{1}{3}},
\end{equation}
where $N_{k}$ is the total number of {\bf k}-points and V$_\t{BZ}$ the BZ volume.

To discuss and analyze numerical errors, we performed total-energy calculations  for fixed geometries, i.e., without any  relaxation, using a representative set of numerical settings. These are $\vec{k}$-point densities of 2, 4, and 8~\AA, respectively, and choices of basis sets that are described in detail in the \SM. They reflect settings typically used in production calculations and also include extremely precise numerical settings that ensure convergence in total energy of~$<0.001$~eV/atom. The latter are termed ``fully converged'' reference when we evaluate the error occurring with less precise~(typical) settings. To make sure that no other numerical errors  cloud the ones stemming from the $\vec{k}$-grid and the basis set, all other  computational parameters --for example, the convergence thresholds for self-consistency-- were chosen in an extremely conservative way, as detailed in the \SM. 

To cover the chemical space in these benchmark calculations, a set of representative materials has been chosen.  
This includes the 71 elemental  solids that have been studied in the aforementioned work by  Lejaeghere and coworkers\cite{Cottenier16} and also includes prototypical binary materials (one for each element with atomic number~$\le71$;  noble gases excluded). The  atomic structures and detailed geometries were taken from the experimental Springer materials database\cite{SpringerMaterials} by selecting the energetically most stable binary structure for each particular element.\footnote{We us the $T=0K$ experimental geometries. Zero-point vibrational effects are included in these experimental values and are not corrected for in the calculations. This is fine as we only need a consistent treatment for all calculations and materials.} On top of that, 10 ternary materials were chosen from the NOMAD  Repository.\cite{NomadRepo} A detailed list including space groups, stoichiometric formulae, structures, and references to the original scientific publications is given in the \SM.

In the following, we focus on the convergence and related errors of two fundamental properties,~i.e.,~the absolute total energies~$E_\t{tot}$ and relative energies~$E_\t{rel}$. The latter were computed as the total-energy difference between the original unit cell and an expanded cell,  with 5\% larger volume and scaled internal atomic positions. While $E_\t{tot}$ includes both the energetic contribution from core and valence electrons, $E_\t{rel}$ is less sensitive to contributions from the core and semi-core electrons due to benign error cancellation. Accordingly, $E_\t{rel}$ is a good metric to quantify the  typically needed numerical precision for energy differences as well as potential-energy  surfaces. It also sheds light on the errors that would occur in properties derived from the total energy, like elastic constants, vibrational properties, and alike. In our evaluations, the error for one material~$i$ in a data set $x_i$ is always defined with respect to the ``fully converged'' reference  value~$c_i$, as indicated by the notation $\Delta x_i=x_i-c_i$,~e.g.,~$\Delta E_{\t{tot},i}$ for the total energy error of material~$i$. 
To statistically analyze the errors across the full set of materials with $N$ entries, we report the mean absolute error
\begin{equation}
   \langle \Delta x \rangle =\frac{1}{N}\sum_i^N |\Delta x_i|
\end{equation}
and the maximum error
\begin{equation}
   \t{max}(\Delta x) = \max_i \left\vert \Delta x_i \right\vert .
\end{equation}
Here, we limit the discussion to data computed with the PBE xc-functional. The 
numerical errors occurring with a different  type of generalized gradient approximation (GGA) 
or the local-density approximation (LDA) show the same qualitative behavior and only minor 
quantitative differences (see \SM).
However, quantitative differences occur for beyond-DFT methods, as discussed in Sec.~\ref{Outlook}.

\section{Results}
In the following,  we first summarize the trends observed for the elemental solids (Sec.~\ref{elemental}). When discussing errors related to the basis set, we always compare to calculations that are ``fully converged'' with respect to $\vec{k}$-points.
Likewise, errors arising from an insufficient $\vec{k}$-point density are discussed for ``fully converged'' basis sets, since the  errors arising from either source can be considered independent of each other.  In all cases, a simple summation approach with a Fermi-function smearing of 100~meV is used for the BZ integration. The observed trends allow us to propose a simple, but versatile mathematical model to estimate the error associated to the basis set for \emph{any} compound and \emph{any} of the investigated codes, as exemplified in Sec.~\ref{BinTern} for binary and ternary materials.

The results of this work are available for further analysis,  both as raw data in the NOMAD Repository\cite{NomadRepo} and as a Jupyter notebook  in the NOMAD Data-Analytics Toolkit~(\url{https://analytics-toolkit.nomad-coe.eu/tutorial-error-estimates}). Therein, errors for arbitrary systems can be calculated via an easy-to-use interface for various numerical settings for \exciting, \FHIaims, and \GPAW. The  corresponding Python code can be modified and extended for custom purposes.

\subsection{Elemental Solids}
\label{elemental}
First, we address the convergence with respect to the size viz. degree of completeness of the basis set.  The results are shown in Fig. \ref{fig:error_basis}. In the case of \exciting, 
the  atom-specific settings, which are kept fixed in all calculations, correspond to a sizable number of local orbitals that ensure well-converged ground-state calculations and transferability between different compounds.  The remaining (and most widely used) parameter to judge the quality of the  plane-wave basis is $RK_\t{max}$, which is the product of the radius of the smallest atomic sphere and the plane-wave cutoff (for details, see Ref.~\onlinecite{Gulans2014exciting}). Choosing the {\it optimal} value $RK_\t{max}^\t{opt}$ such that it corresponds to a convergence of the total energy of about 0.1~meV/atom (for details, see the \SM), we use the squared fraction~$(RK_\t{max}/RK_\t{max}^\t{opt})^2$ to label the basis set quality, 
see \SM\ for details.
For \FHIaims, which uses tabulated, 
chemical-species-specific sets of NAOs, the number of NAOs per electron is used as  metric. Note that these NAOs come in \textit{tiers} that group different angular momenta.\cite{Blum09} The average number of basis functions per electron 
present in these tiers and in the species-specific suggested settings (``light'', ``tight'') provided by the  \FHIaims\ developers are shown as black and gray vertical lines  in the figures. Since the ``translation'' from the number of NAOs into this metric requires binning (not all elemental solids appear for all values of the x-axis), the reported errors do not decrease monotonically. It is important to note that tier 4 sets are not provided for all elements, but only for those species for which such an additional set of basis functions improved the description of the electronic structure during the basis-set construction procedure.\cite{Blum09} Accordingly, only these problematic elements determine the errors shown for 9 and more NAOs per electron. The more benign elements, that are already fully converged in this limit, no longer enter the shown average error, since the developer-suggested settings do not allow for more than 8 NAOs per electrons for these species.  
In the plane-wave code \GPAW\ the basis set is characterized by the cut-off energy $E_\t{cut}$,  i.e., all plane waves with a kinetic energy smaller than $E_{\rm cut}$ are included in the basis set. Note
that this affects the convergence of relative energies, since, for the same value of $E_\t{cut}$, cells with different volume contain different number of plane waves.

As evident from Fig.~\ref{fig:error_basis}, the errors in the total energy exhibit a systematic convergence with increasing basis-set size for all three codes.
Generally, the maximum error in the total energy can be even roughly one order of magnitude larger than the average error.
This is due to the fact that numerical errors are element specific,~i.e.,~some chemical species require a large basis set to
be described precisely. This is reflected by the fact that the difference between average and maximum error 
is more pronounced in the  results for \FHIaims\ and \GPAW~(Fig. \ref{fig:error_basis}) due to the metric chosen to quantify the basis-set completeness,~i.e.,~the $x$-axis in this figure. While \FHIaims\ and \GPAW\ use an absolute metric, \exciting\ uses a relative one,~i.e., fractions of species-specific values~$RK_\t{max}^\t{opt}$. In this case, the fact that the developers provide well-balanced, species-specific values for $RK_\t{max}^\t{opt}$ ensures that a similar precision is achieved for all species at a specific fraction
of $(RK_\t{max}/RK_\t{max}^\t{opt})^2$. In turn, this leads to a more consistent precision across material space and thus to smaller maximum errors at a given value of $(RK_\t{max}/RK_\t{max}^\t{opt})^2$. 
For all three codes, the average and maximum errors in total energies are roughly one to two orders of magnitudes larger than the ones for relative energies.  Again, this finding reflects that the main source for imprecisions 
in the total energy is species specific and leads to a beneficial error cancellation in energy differences. 

Eventually, it is important to note that the numerical errors vary considerably for different species, types of bonding, and across methodologies, as detailed in the Supp. Material. 
Naturally, plane waves are more suitable for quasi-free-electron systems like aluminum, whereas NAOs perform better for inert elements like rare gases or localized covalent bonds. 
These observations, dating back to the early days of electronic-structure theory and predating modern DFT implementations, are among the historical reasons\cite{Koelling:1981dd} 
that actually led to the development of the different methodologies discussed in this paper. Accordingly, also the above described trends for the numerical errors, their influence
on computed observables, and their numerical as well physical origin have been discussed\cite{Zunger:1979hr,Weinert:1982kb,Holzschuh:1983gr} and reviewed\cite{Devreese:1985ip} in literature before. Most importantly, the finding that errors are 
largely species-specific can be rationalized by the fact that changes in the kinetic energy of core electrons, despite being orders of magnitude larger than total-energy changes, vanish to first order in charge-density differences.\cite{VonBarth:1980jv}
For instance, this aspect is directly exploited in the \VASP\ code\cite{Kresse:1996kg,Kresse:1996vf}
for an automatic convergence correction\cite{Rappe:1990dk}\footnote{Due to this automatic convergence correction, the total energy output of VASP does not necessarily decrease monotonically when $E_\t{cut}$ is 
increased, as it is the case in most common PAW implementations. Accordingly, an analysis of this code-specific aspect goes beyond the scope of this paper. Nonetheless, a complete, consistent VASP dataset covering the materials discussed in this work is available via the NOMAD repository at \href{https://dx.doi.org/10.17172/NOMAD/2020.07.29-1}{DOI:10.17172/NOMAD/2020.07.29-1}. }.
In Sec.~\ref{BinTern}, we will exploit this fact for the three codes \exciting, \FHIaims, and \GPAW\ to predict errors \textit{a priori} for multi-component systems using information from the elemental solids.

\begin{figure}[b!]
  \centering
  \includegraphics[clip,width=0.9\linewidth]{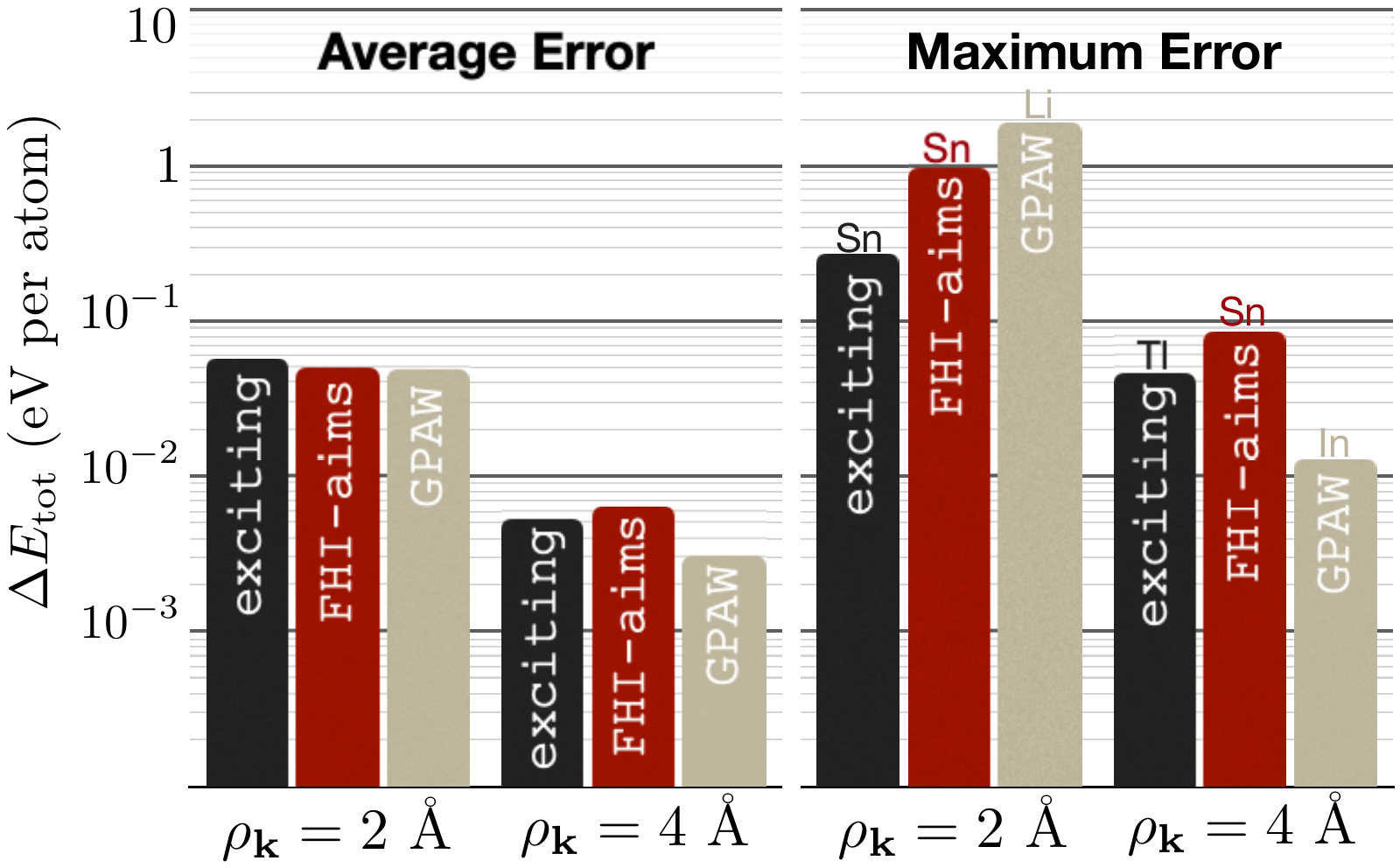}
  \caption{ Average (left) and maximum error (right) observed for the elemental solids for two different $\mathbf{k}$-point densities $\rho_\vec{k}$ with \exciting, \FHIaims, and \GPAW. In all three codes, the calculations were carried out using the PBE xc-functional. A simple summation with a Fermi-function smearing of 100~meV is used for the BZ integration to facilitate comparison between codes.}
\label{fig:error_k-mesh}
\end{figure}

\begin{figure*}
  \centering
  \includegraphics[width=0.95\linewidth]{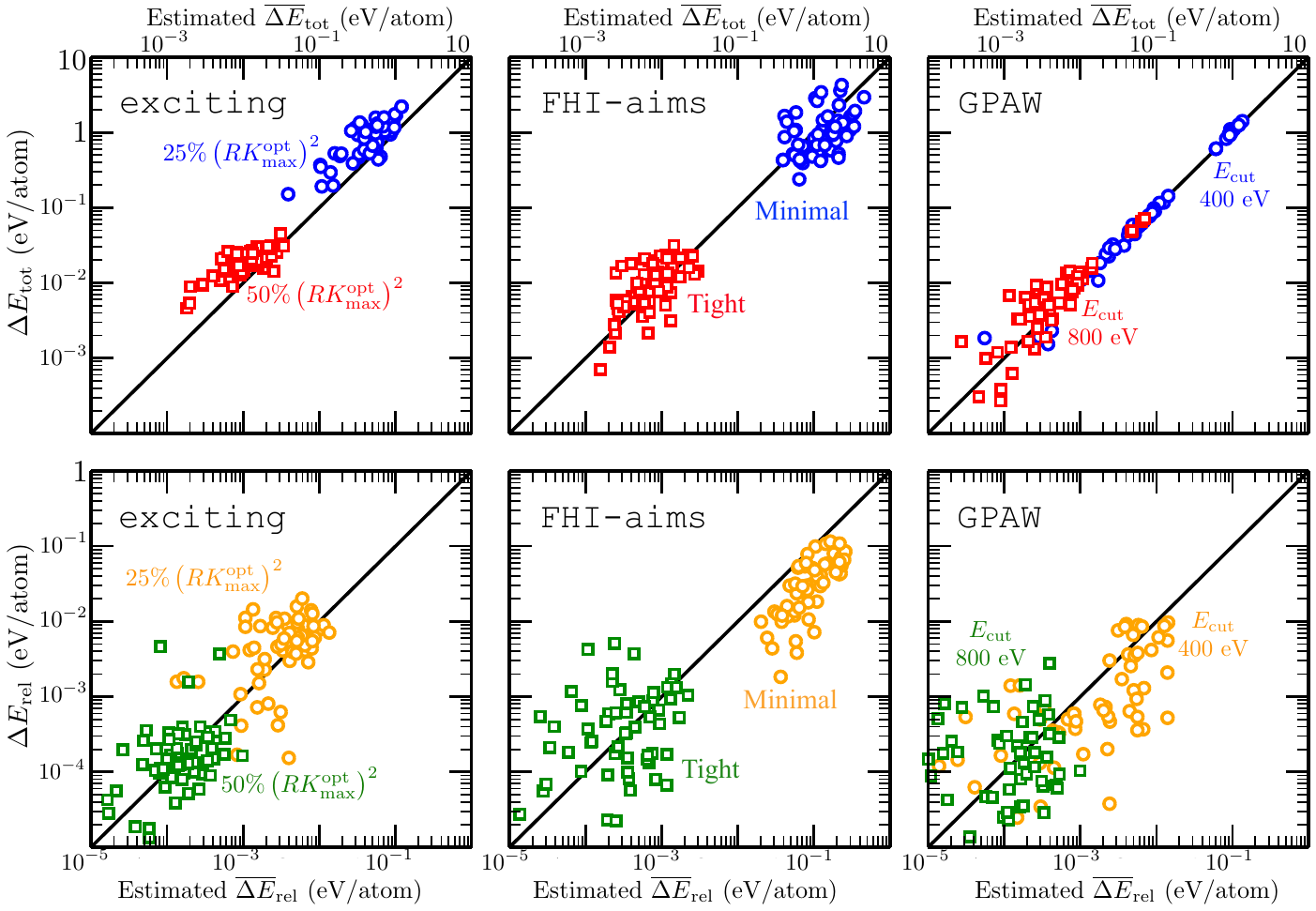}
  \caption{Estimated vs. actual numerical errors in $E_\text{tot}$~(upper row) and $E_\text{rel}$~(lower row) for 63 binary systems, considering two 
basis-set sizes for each of the three employed codes. The structures were chosen from the experimental Springer materials database\cite{SpringerMaterials} by selecting the energetically most stable binary structure for each particular element. Note the logarithmic scales and the different energy windows in the upper and lower row.}
\label{fig:binaries}
\end{figure*}

Let us now inspect the errors in total energies that arise due to the finite reciprocal-space grid. Figure~\ref{fig:error_k-mesh} shows results for $\vec{k}$-point densities of 2~\AA\ and 4~\AA. Data obtained with a $\vec{k}$-point density of 8~\AA\ serves as ``fully converged'' reference. The rather large observed errors result from the fact that many elemental solids are metallic with a more involved shape of the Fermi-surface, so that a substantial number of $\vec{k}$-points is required to reach convergence. Quite consistently, all codes yield average errors of the same order of magnitude if the same $\vec{k}$-point densities are used, despite the fact that the three codes handle the numerical details of the  reciprocal-space integration differently. This is reflected in the maximum errors, which vary slightly more between codes than the average ones.
Again, we observe that the maximum error is approximately one order of magnitude larger than the average error.

\begin{figure*}
  \centering
  \includegraphics[clip,width=1.9\columnwidth]{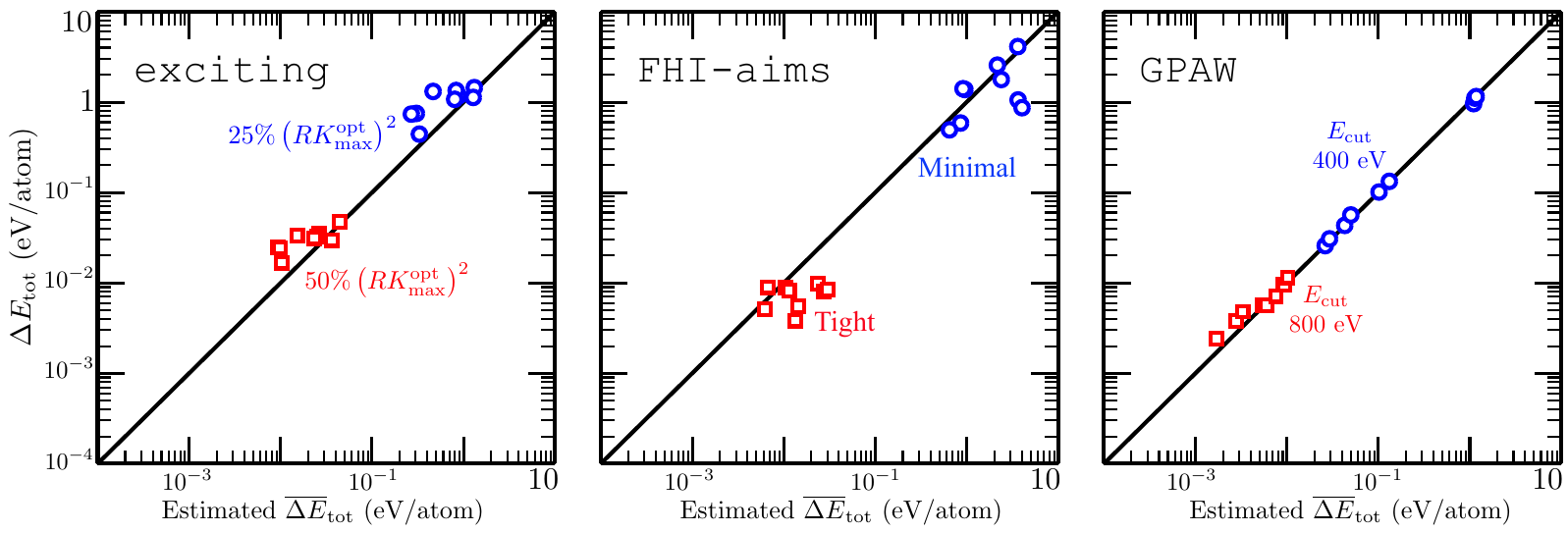}
  \caption{ Estimated vs. actual numerical error in the total energy of ternary systems, using two basis-set sizes in each of the employed codes. Note that 
Ca$_2$CdP, Na$_2$IrZn, and Li$_2$SnCl exhibit very similar errors and are thus not fully visible for all codes and levels of precision.}
\label{fig:ternaries}
\end{figure*}

\subsection{Predicting Errors for Binary and Ternary Systems}
\label{BinTern}
Following our discussion of the  errors in total and relative energies of elemental solids stemming from the basis-set incompleteness, we propose to estimate the  corresponding errors for multicomponent systems by linearly combining the respective errors observed for the constituents in the elemental-solids calculations at the same settings. This follows the above discussed observation that there are chemical species that require larger basis sets to reach convergence.
This is in fact independent of the employed code.  For the error in the total energy,\footnote{In the case of O, F, and N, the elemental solid is a molecular crystal that is not a good representative for the binding in oxides, fluorides, and nitrides. For this reason, we determined the values of~$\Delta E$ for these particular elements from the binaries MgO, NaF, and BN by inverting Eq.~(\ref{eq:prederr})}  we simply assume: 
\begin{equation}
{\overline {\Delta E}_\t{tot}}=\frac{1}{N}\sum_I N_{I}\Delta E_{\t{tot},I}
\label{eq:prederr}
\end{equation}
$N_I$ being the number of atoms of species $I$. For ${\overline {\Delta E}}_\t{rel}$ we proceed analogously. 

To validate the ansatz of Eq.~(\ref{eq:prederr}), we have computed the total and relative energy errors for 63 binary solids using the exact same strategies used
for the elemental solids in Sec.~\ref{elemental}. In Fig.~\ref{fig:binaries}, we then compare these real errors observed in the calculations for binary systems for two basis-set sizes for each of the three codes to the estimated errors obtained via Eq.~(\ref{eq:prederr}). As shown in these plots, we generally obtain
quite reliable total energy predictions for all three codes by this means. For the  total energies (top panels), we observe better predictions when an ``unbiased'' and smooth metric is used to characterize the basis-set completeness. For instance, \GPAW, which uses the atom-independent plane-wave cutoff $E_\t{cut}$, 
yields an almost perfect correlation between predicted and actual total energy errors. Conversely, 
more scattering is observed for \FHIaims, which uses an atom-specific, granular metric with different NAOs for each atom. 
Nonetheless, we find a clear correlation between the predicted, ${\overline {\Delta E}_\t{tot}}$, and the actual errors, ${\Delta}E_\t{tot}$, for all codes.
In particular, this holds for absolute energy errors larger than $>10$~meV/atom. 
This demonstrates that the relatively intuitive relation  formulated in Eq.~(\ref{eq:prederr}) can serve as a reliable estimate for the error associated with a particular total-energy calculation. 

For the relative-energy errors shown in the lower half of Fig.~\ref{fig:binaries}, we observe more scattering and a less neat correlation between predicted and actual errors. The reason for that is twofold: First, benign error cancellation reduces
numerical errors in relative energies, since total energy differences are inspected. In other words, a large portion of the species-specific errors described by Eq.~(\ref{eq:prederr}) cancel each other out when computing relative energies as a difference. For this exact reason, relative energies are generally less affected 
by numerical errors~(see Fig.~\ref{fig:error_basis} and its discussion). 
Second, relative errors are --in contrast to total energies-- non-variational,~i.e.,~they do not necessarily 
decrease monotonically with basis-set size. The reason is that the errors associated to the two total energies
entering the relative energies typically do not decrease at the exact same rate.
Still, the relative energy error estimates for all codes are reliable enough in the respective energy window of interest,
hence allowing us to compare relative energies obtained from different codes with different settings. 

The data shown and discussed for the binary materials suggest that Eq.~(\ref{eq:prederr}) can be used to estimate the total energy errors for \emph{any} multi-component system. As an example, we demonstrate this in Fig.~\ref{fig:ternaries}, in which the same comparison between predicted and actual total energy errors is made for ten ternary systems, which were selected from the huge pool of  compounds available in the NOMAD Repository\cite{NomadRepo} so to cover material and structural space. Also in this case, the same quantitative and qualitative behavior as discussed for Fig.~\ref{fig:binaries} is observed. The relatively simple approach of Eq.~(\ref{eq:prederr}) is
able to correctly predict the numerical errors also in these ternary systems. This further substantiates that the described approach is not only applicable to the relatively simple binary systems discussed in Fig.~\ref{fig:binaries}, but also to more complex systems, as the ones found in electronic-structure materials databases.

\section{Outlook}
\label{Outlook}

\begin{figure}[b]
  \centering
  \includegraphics[clip,width=0.85\columnwidth]{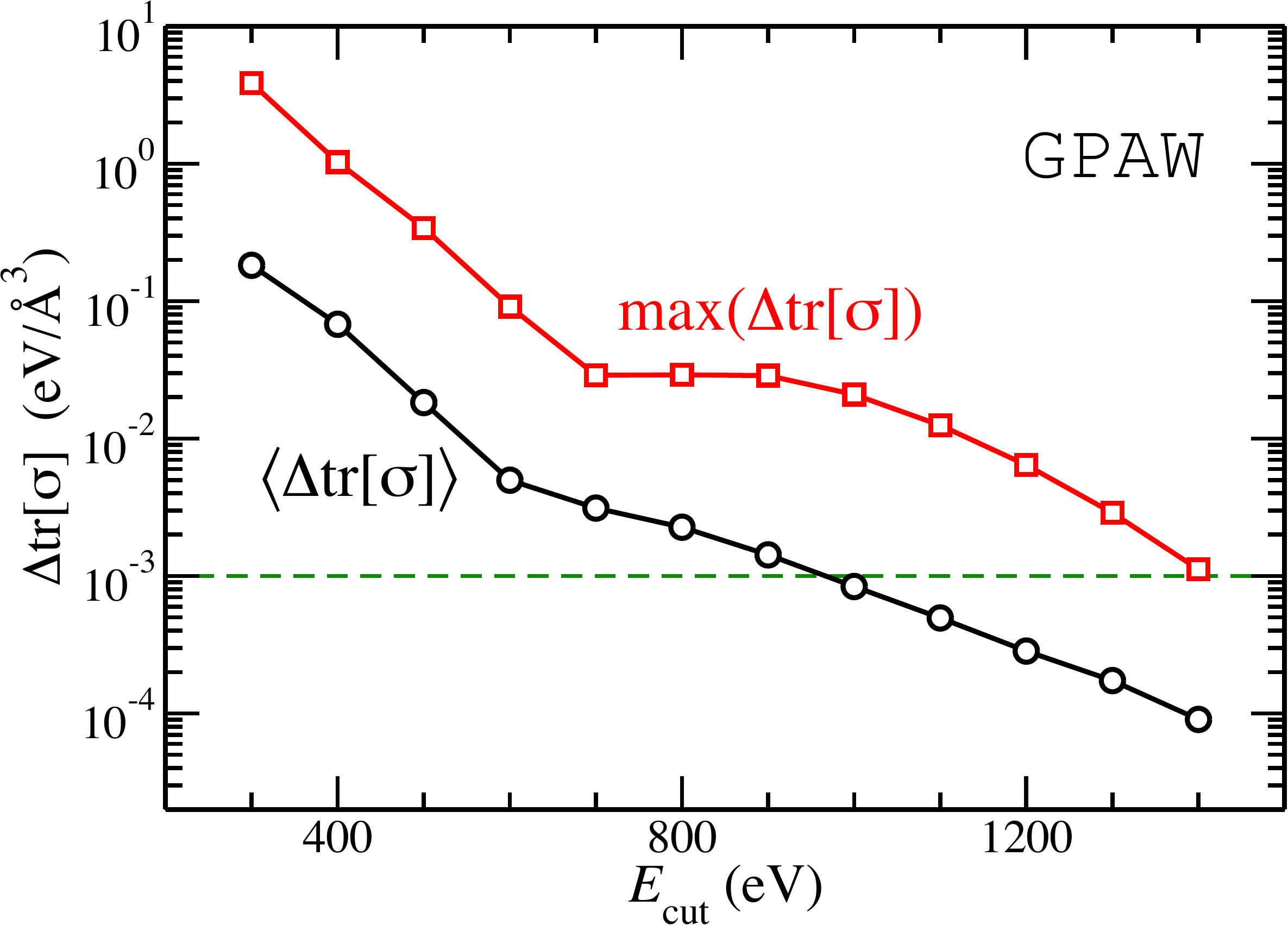}
  \caption{Average~(black circles) and maximum numerical error~(red squares) for the trace of the stress tensor~$\mathbf{\sigma}$ as function of the cut-off energy~$E_\t{cut}$
  for 71 elemental solids when the experimental lattice constant is used. All calculations were performed  with \GPAW\ using the PBE functional and a 8~$\mbox{\AA}$~$\vec{k}$-point density. The dashed green
  line highlights the typical error acceptable in high-accuracy lattice relaxations.}
\label{fig:stress_GPAW}
\end{figure}

The focus of the formalism presented in this work lays on the analysis of total and relative energies, since those are the most fundamental
quantities produced in electronic-structure-theory calculations. However, such first-principles approaches also allow computing many other 
material properties, ranging from structural parameters, over thermodynamic expectation values, to electronic properties. Generally, these 
quantities will exhibit a different convergence behaviour than the total and relative energies. In particular, this is the case for non-variational 
properties that do not depend monotonically on the basis-set size and $\textbf{k}$-grid density.

\begin{figure}[b]
  \centering
  \includegraphics[clip,width=0.85\columnwidth]{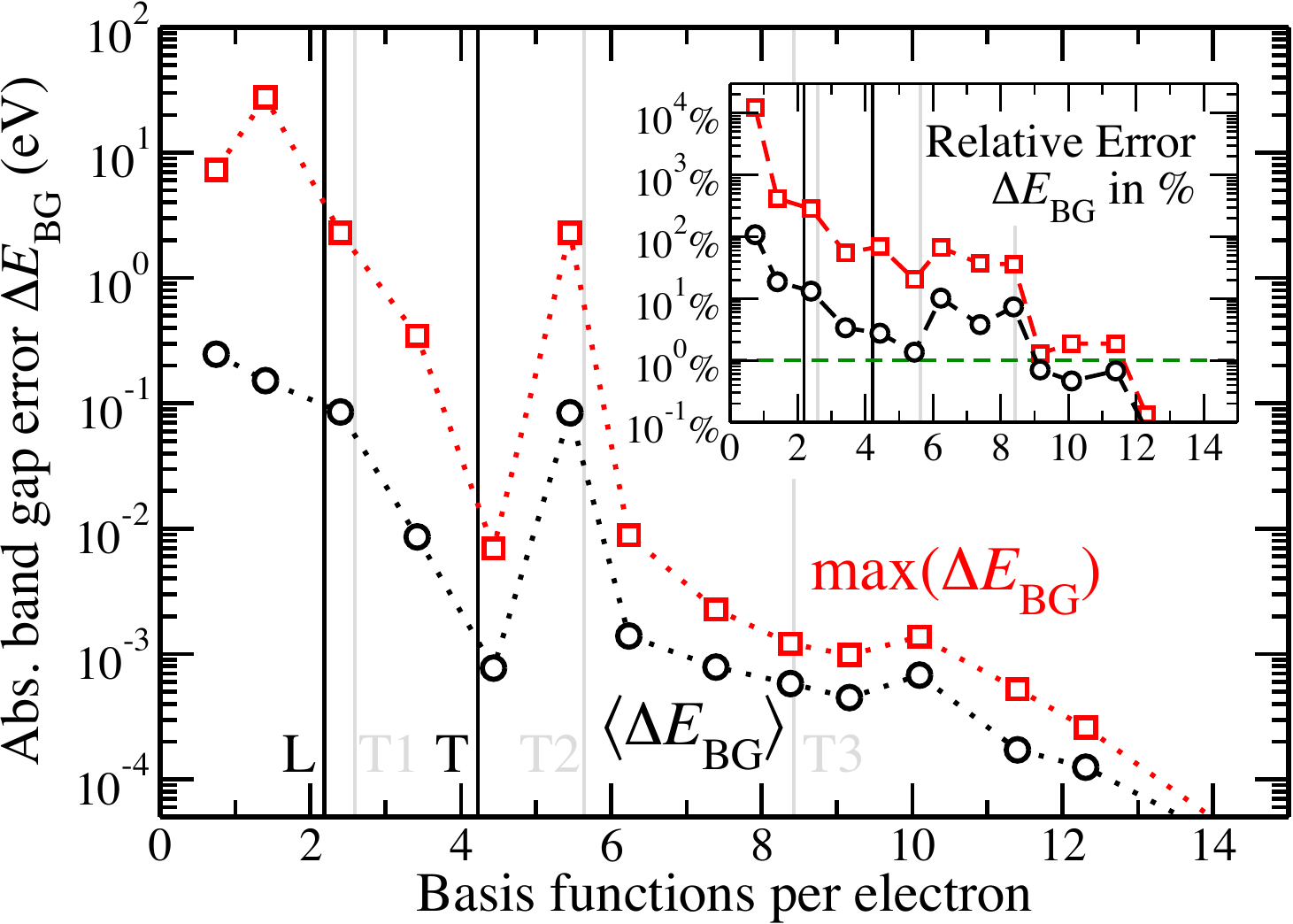}
  \caption{Average~(black circles) and maximum numerical error~(red squares) for the Kohn-Sham band gap~$E_{\t{BG}}$ 
  as function of the number of basis functions per electron for 71 elemental solids. Vertical black~(L/T) and gray lines~(T1,T2,T3) denote the average number of basis functions for different settings and tiers,
  as in Fig.~\ref{fig:error_basis}. All calculations were performed  with \FHIaims\ using the PBE functional and an 8~$\mbox{\AA}$~$\vec{k}$-point density. The inlet shows
  the relative errors in percent with respect to the fully converged band gap. The dashed green line highlights the typical error acceptable for obtaining reliable electronic properties,~e.g.,~Fermi-energies
  and effective masses.}
\label{fig:gap_FHIaims}
\end{figure}

As an example, we discuss the numerical errors associated with the evaluation of the stress tensor~$\mathbf{\sigma}$. Its components are defined as~\cite{Nielsen:1983ti,Knuth:2015kc}
\begin{equation}
\sigma_{\lambda \mu}=\left.\frac{1}{V} \frac{\partial E_{\mathrm{tot}}}{\partial \varepsilon_{\lambda \mu}}\right|_{\varepsilon=0} \;,
\label{eq:stress}
\end{equation}
i.e., as the total energy derivatives with respect to symmetric strain deformations~$\varepsilon_{\lambda \mu}$ for the Cartesian axes~$\lambda,\nu$
normalized by the unit-cell volume~$V$.
Despite the fact the stress is defined as a total energy derivative, it is well known~\cite{Bernasconi} that it is particularly sensitive to the value 
of $E_\t{cut}$ chosen in plane-wave calculations. This is further demonstrated for the \GPAW\ code in Fig.~\ref{fig:stress_GPAW} using the trace of the
stress tensor~$\t{tr}\left[\mathbf{\sigma}\right]$, as computed for the experimental lattice constants and structures. Qualitatively, the average and 
maximum errors observed for the stress resemble the behaviour observed for \GPAW's total energy convergence quite closely, as a comparison of Figs.~\ref{fig:error_basis} 
and~\ref{fig:stress_GPAW} reveals. This is not too surprising, given that stress and total energy are directly related via Eq.~(\ref{eq:stress}). However,
obtaining meaningful values for the stress,~i.e.,~values accurate enough to perform reliable structure relaxations, requires roughly $50$\% higher cut-off
energies~$E_\t{cut}$ than needed to obtain reasonably converged total energies. Let us note that the contributions to the numerical error in 
the stress tensor stemming from the finite $\mathbf{k}$-grid density are much smaller than those arising from~$E_\t{cut}$, as found for the total energy
before~(see Fig.~\ref{fig:error_k-mesh}). With respect to the basis-set convergence, the observed trends suggest that the strategy devised in this work for 
total and relative energies might also be useful for estimating errors in first-order derivatives of the total energy,~i.e.,~for forces and stresses, which only depend on an accurate description
of occupied electronic states.~\cite{Gonze:1989vv} More data, especially for structures far from equilibrium, is needed to further investigate this
hypothesis and to develop accurate error-estimate models for such quantities.

\begin{figure}[t]
  \centering
  \includegraphics[clip,width=0.85\columnwidth]{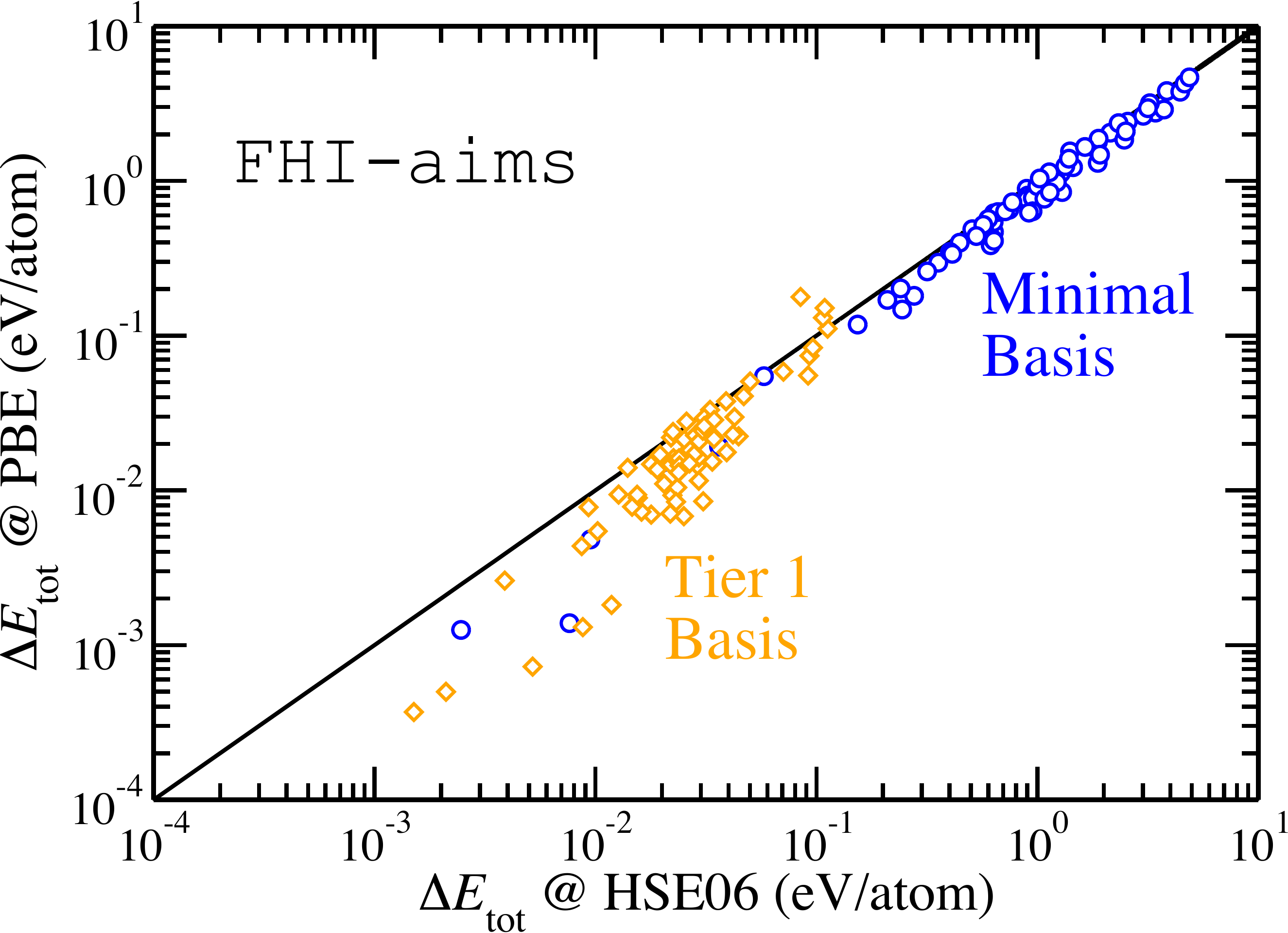}
  \caption{Correlation of the numerical errors obtained for the total energy of the 71 elemental solids for different basis sets~(minimal, ``Tier 1'')  
  for the semi-local PBE and the generalized hybrid HSE06 functional. All calculations are performed with {\FHIaims}, ``light'' numerical
  settings, and a 2~$\mbox{\AA}$~$\mathbf{k}$-point density.}
\label{fig:HSE06}
\end{figure}

Not all material properties of interest solely depend on occupied electronic states,~e.g,~evaluating opto-electronic properties specifically requires 
the eigenvalues (and/or wave functions) of unoccupied electronic states. As an example for such kind of properties, we show in Fig.~\ref{fig:gap_FHIaims} the 
error of the Kohn-Sham band gap,~$\Delta E_{\t{BG}}$, 
as obtained for the 71 elemental solids from band-structure calculations with \FHIaims\ along high-symmetry paths in the Brillouin zone.~\cite{Setyawan:2010hl} 
The comparison of Fig.~\ref{fig:gap_FHIaims} with the respective total energy convergence plot in Fig.~\ref{fig:error_basis} shows that the range observed for both 
average and maximum errors in $E_{\t{BG}}$ spans almost twice the orders of magnitude obtained for $E_{\t{tot}}$, substantiating that larger basis sets are
required to converge $E_{\t{BG}}$. Furthermore, we note that the numerical errors do not decrease monotonically. In part, this is a consequence of the fact
that the band-gap is a difference of two values that exhibit different, non-variational convergence.
Furthermore, we see again the effect of  the employed ``binning'' procedure discussed for $E_{\t{tot}}$ above~(e.g.~the peak at 5.5 basis functions per electron).
In the case of the band gap, the latter is particularly important, since the calculated band gaps span a wide range, starting from virtually zero,~e.g.,~for graphite,
and reaching 17~eV for the rare gas helium. For this exact reason, the relative numerical errors for $E_{\t{BG}}$, shown in percent of the converged value in the 
inlet of Fig.~\ref{fig:gap_FHIaims}, exhibit a more regular -- but still non-monotonic -- behaviour. As it was observed for the evaluation of the stress, computing 
reasonably converged band-gaps hence requires roughly 50\% larger basis sets than needed to achieve total energy convergence.

As noted in the introduction, we have restricted our analysis to (semi-)local xc-functionals, since such kind of calculations
are the current workhorse in computational \textit{high-throughput} studies and hence constitute the uttermost majority of data stored in existing electronic-structure
theory databases.\cite{Draxl:2018ec} However, it is well known that beyond-DFT methods require larger basis sets to achieve convergence in total energy.\cite{DFTBasisSet} 
For the generalized hybrid functional HSE06,\cite{Krukau:2006jq} which incorporates a fraction of non-local, exact exchange, this is demonstrated in Fig.~\ref{fig:HSE06},
which shows the correlation between the numerical errors observed in the total energy of the 71 elemental solids for the PBE and HSE06 functional, respectively. Especially
when compared to the LDA/PBE correlation plot shown in the \SM, it is obvious that the numerical errors are typically larger in HSE06 calculations. 
Nonetheless, there is a clear qualitative correlation between PBE and HSE06 errors, suggesting that the strategies developed in this work might also be useful for
beyond-DFT databases.

\section{Conclusions}
In this study, we 
presented an extensive, curated data set obtained by three conceptually very different electronic-structure methods. This set contains elemental solids, binary, and ternary materials for various combinations of computational parameters. The data have been used to understand and predict the errors of calculations with respect to the basis-set quality. More specifically, we have shown that the errors for arbitrary systems can be estimated from the errors obtained from systematic calculations for related elemental solids, as exemplified for 63 binaries and 10 ternary systems covering 13 different
space groups. Let us emphasize that the presented findings are not code-specific,~i.e.,~limited to {\exciting}, {\FHIaims}, and {\GPAW}. Rather, the qualitative trends 
observed for the  \textit{linearized augmented plane-waves plus local orbitals}, the \textit{linear combination of numeric atom-centered orbitals}, and the
\textit{projector-augmented wave} formalisms, respectively, generally hold for all implementations of these approaches, thus covering the vast majority of codes 
present in current material databases. Obviously, quantitative error estimates for individual codes depend on the details of the implementations and 
basis sets,~e.g.,~the chosen local orbitals, the exact definition of the NAOs, or the employed PAW potentials, and thus require code-specific reference 
calculations for the elemental solids.
That given, the developed formalism, which gives surprisingly good results for total energies despite its conceptual simplicity, can be incorporated into computational materials databases to estimate errors of stored data. This is a prerequisite for operating on data collections that originate from different computations, performed with different computer codes and/or different precision. Our work may serve as a starting point for more sophisticated concepts to quantify numerical errors and uncertainties, especially for more complex materials properties
that do not necessarily depend monotonically on the basis-set size,~e.g.,~band gaps, forces, vibrational frequencies, and the relative energies discussed in this work.

\section{Data Availability}
All presented data,~i.e.,~in- 
and output files for all electronic-structure theory codes, is available at the NOMAD Repository\cite{NomadRepo} for \href{https://dx.doi.org/10.17172/NOMAD/2020.07.15-1}{\exciting}, \href{https://dx.doi.org/10.17172/NOMAD/2020.07.27-1}{\FHIaims}, \href{https://dx.doi.org/10.17172/NOMAD/2020.08.20-1}{\GPAW}, as well as \href{https://dx.doi.org/10.17172/NOMAD/2020.07.29-1}{\VASP}. Additionally, the data
can be explored interactively using the Jupyter notebook in the NOMAD Data-Analytics Toolkit~(\url{https://analytics-toolkit.nomad-coe.eu/tutorial-error-estimates}).

\section{Author Contributions}
CC, LMG, and MSc designed the database-creation protocol; BB and MSt developed the ASE-based scripts for setting up and performing the calculations. 
SL and AG performed the \exciting, BB and CC the \FHIaims, MSt and JJM the \GPAW, and EW and OTH the \VASP~calculations. 
BB, MSt, SL, LMG and CC wrote the notebook to evaluate and analyze the data. CD, KST, and MSc
ideated the project that was led and coordinated by CC. All authors contributed to the discussion of the results and to the writing of the manuscript.

The authors declare that there are no competing interests.

\begin{acknowledgments}
This project has received funding from the European Union’s Horizon 2020 research and innovation program under grant agreement No. 676580 and No. 740233~(TEC1p). OTH and EW gratefully acknowledge funding by the Austrian Science Fund, FWF, under the project P27868-N36.
We gratefully acknowledge the help from Mohammad-Yasin Arif and Luigi Sbailò for producing the final version of the Jupyter notebook and publishing it on the NOMAD AI toolkit.
\end{acknowledgments}

\bibliography{error_bars}
\end{document}